\documentclass[pdflatex,sn-mathphys-num]{sn-jnl}

\usepackage{graphicx}%
\usepackage{multirow}%
\usepackage{amsmath,amssymb,amsfonts}%
\usepackage{amsthm}%
\usepackage{mathrsfs}%
\usepackage[title]{appendix}%
\usepackage{xcolor}%
\usepackage{textcomp}%
\usepackage{manyfoot}%
\usepackage{booktabs}%
\usepackage{algorithm}%
\usepackage{algorithmicx}%
\usepackage{algpseudocode}%
\usepackage{listings}%
\usepackage{booktabs}  
\usepackage{multicol}
\usepackage{multirow}
\usepackage{placeins} 
\usepackage{graphicx,bm}
\usepackage{lineno,hyperref}
\theoremstyle{thmstyleone}

\theoremstyle{thmstyletwo}

\theoremstyle{thmstylethree}%

\begin{document}

\title[A Counterfactual Approach for Addressing Individual  User Unfairness in Collaborative Recommender System]{A Counterfactual Approach for Addressing Individual  User Unfairness in Collaborative Recommender System}

\author*[1]{\fnm{Nikita} \sur{Baidya}}\email{518cs1012@nitrkl.ac.in}

\author[2]{\fnm{Bidyut} \spfx{Kr.} \sur{Patra}}\email{bidyut.cse@itbhu.ac.in}

\author[1]{\fnm{Ratnakar} \sur{Dash}}\email{ratnakar@nitrkl.ac.in}

\affil*[1]{Pattern Recognition Laboratory, \orgdiv{Department of Computer Science and Engineering}, \orgname{National Institute of Technology Rourkela}, \orgaddress{\state{Odisha}, \country{India}}}

\affil[2]{\orgdiv{Department of Computer Science and Engineering}, \orgname{Indian Institute of Technology (BHU) Varanasi}, \orgaddress{\state{Uttar Pradesh}, \country{India}}}

\abstract{Recommender Systems (RSs) are exploited by various business enterprises to suggest their products (items) to consumers (users). Collaborative filtering (CF) is a widely used variant of RSs which learns hidden patterns from user-item interactions for recommending items to users. Recommendations provided by the traditional CF models are often biased. Generally, such models learn and update embeddings for all the users, thereby overlooking the biases toward each under-served users individually. This leads to certain users receiving poorer recommendations than the rest. Such unfair treatment toward users incur loss to the business houses. There is limited research which addressed individual user unfairness problem (IUUP). Existing literature employed explicit exploration-based multi-armed bandits, individual user unfairness metric, and explanation score to address this issue. Although, these works elucidate and identify the underlying individual user unfairness, however, they do not provide solutions for it.   

In this paper, we propose a dual-step approach which identifies and mitigates IUUP in recommendations. In the proposed work, we counterfactually introduce new interactions to the candidate users (one at a time) and subsequently analyze the benefit from this perturbation. This improves the user engagement with other users and items. Thus, the model can learn effective embeddings across the users.
To showcase the effectiveness of the proposed counterfactual methodology, we conducted experiments on MovieLens-100K, Amazon Beauty and MovieLens-1M datasets. The experimental results validate the superiority of the proposed approach over the existing techniques.}


\maketitle

\section{Introduction}\label{sec1}
Digital platforms have gained immense popularity  and have become an indispensable part of our daily lives \cite{impact, role, factors}. Whether we are shopping from e-commerce websites or watching content on streaming platforms or discovering new friends on social media, these digital hubs have made our lives convenient and seamless \cite{let, influence}. One of the key challenges these applications face is that they are overloaded with information (items). This may cause decision fatigue and can overwhelm a user \cite{bib1}. To help narrow-down the overwhelming amount of choices for a user, these sites utilize recommender systems (RSs). RSs employ artificial intelligence techniques for offering personalized content, product, and services to a user \cite{bib2, pricai, luoneural}. 

Based on the information utilized for generating recommendations, RSs can be broadly categorized into content-based filtering (CBF), collaborative filtering (CF) and hybrid \cite{bib3, jannachrecommender, adtoward, ricciintro}. Hybrid RSs combine the recommendation strategies of CF and CBF models for generating recommendations. CBF models are trained on item metadata (features), whereas, CF models learn hidden patterns (latent features or embeddings) from user-item interactions. A major challenge for CBF is that it requires rich and structured information of every item. CF is a more popular and widely used RS than CBF as it overcomes this challenge and is scalable in real-world recommendation environments. Some of the well-known state-of-the-art CF RSs are Matrix Factorization (MF) \cite{bib4}, Neural Matrix Factorization (NMF) \cite{bib5} and Light Graph Convolution Network (LGN) \cite{bib6}. 

Recent studies have revealed that the recommendations provided by RSs are often biased (\cite{bib7,bib8}). The RSs tend to give poorer recommendations to certain users as compared to others \cite{bib9}. As suggested by Li et.al. \cite{bib10}, the premium (paid) users should get superior recommendations however not at the cost of neglecting other users by giving them inferior recommendations. Otherwise, the under-served users will eventually loose interest from the platform, thereby impacting the platform's revenue. Hence, user unfairness is an important challenge that needs to be addressed. We synonymously refer to the term unfairness as disparity throughout this work. There is limited research which examines the interconnection between individual user disparity and user-item interactions. 

Recently, $ACFR$-User$_u$ \cite{bib11} was proposed to address individual user disparity. In this framework, predicted ratings obtained from training DeepFM \cite{bib12} were counterfactually imputed to individual users and pre-trained SVD++ \cite{bib13, bib14} model was trained by minimizing the deviation in predicted and actual ratings for the individual user from the population mean \cite{chengweb} . Their architecture elucidated the users which positively and/or negatively impacted due to addition of new ratings, however they did not provide solutions to mitigate the problem. 

Traditional CF models tend to embed and amplify existing biases in the data \cite{bib15, bib16}. Hence, certain users get worse recommendations than others \cite{bib17}. To address this challenge, we propose a framework which counterfactually adds edges to a user's interaction graph and observes if performing this task leads to any sort of performance gain for the individual user. Additionally, the model also identify users who may contribute in performance gain of the overall users. The central objective of our approach is to identify the individual under-served users who may benefit from adding some extra interaction edges with unobserved items and at the same time may contribute towards the performance improvement for the overall users. The proposed approach specifically targets the uneven distribution of recommendation, seeking to first identify and then improve the recommendation quality for the candidate users and the recommendation quality towards the overall users.

In summary, the contributions of this paper are as follows:
\begin{itemize}
    \item We propose a counterfactual imputation framework to address individual user disparity.
    \item We build an architecture which identifies the under-served users who may benefit from imputation and may contribute to improvement in recommendation for overall users.
    \item We conduct extensive experiments on three datasets to validate the effectiveness of our proposed framework.
\end{itemize}
We proceed our discussion starting with the related work in the next section.

\section{Related Work} 
\label{S:2}
Research in the area of fairness in recommender systems has garnered significance over the recent years \cite{boratto, kheya, luyencontext}. There is a growing emphasis on ensuring comparable treatment for users. In this section, we summarize recent related literature which addresses user disparity. 

Rastegarpanah et al. \cite{bib9} addressed user unfairness by adding fake users. These synthetic users interact with the real items present in the training data. These interactions (ratings) are labeled as antidote ratings. Then, matrix factorization model is trained on the modified training data. This altered training data influences the predictions for the actual users. They additionally utilized individual user unfairness and group user unfairness as loss function to address polarization among all the individual users and the user groups, respectively. 

Li et al. \cite{bib18} categorized the users into two groups based on their number of interactions with items, the total cost of items they have interacted with and the highest cost of item they have purchased. They labeled the users falling in the top 5\% among these categories as advantaged users whereas, the remaining users as disadvantaged users. Following this, they presented a fairness-constrained re-ranking approach which optimized the predictions for disadvantaged users by reducing their performance gap from the advantaged users. This, however resulted in decrease in recommendation quality for the advantaged users. 

Cho et al. \cite{bib19} introduced difference in equal experience across different user groups as a regularization term. The motivation behind this was to make the predictions of recommendation model independent of the user and item groups. Then, they utilized kernel density to estimate these group-conditional probabilities and updated model parameters via gradient descent, thereby reducing group-dependent prediction bias and giving all user groups more equal exposure to diverse group of items.

Do et al. \cite{bib20} addressed individual user disparity based on the notion that a recommender model is fair if it provides recommendation to users based on their own preference rather than based on the preference of some other user. They examined this by treating the task as a pure-exploration multi-armed bandit problem. The users were treated as the arms. Concretely, they introduced an approach which occasionally showed a user another user’s recommendations while enforcing a conservative constraint so overall experience stays close to baseline and does not deteriorate much from its original performance.  

Li et al. \cite{bib11} addressed individual user and item disparity by counterfactually adding predicted ratings onto all the missing interactions of certain randomly selected individual users or items. They exploited individual disparity measures as loss functions to nudge pre-trained recommendation model. They calculated explanation score for each of the candidates (users or items) and suggested that a positive value of the score infer the candidate influenced the model positively and a negative value imply the candidate influenced the model negatively.

To sum up, identification and mitigation of individual user unfairness in recommender systems is an area which needs more exploration. To fill this gap, we propose a framework which leverages a popular graph neural model. We describe the graph model in detail in the next section.

\section{Preliminary}
\label{S:3}
Graph-based RSs are among the most widely utilized CF models due to their superiority in deciphering the intricate user-item relationships. One such RS is LightGCN \cite{bib6} which has gained popularity due to its simplicity. We abbreviate this model as LGCN in this work. In this section, we explain the methodology of this model \cite{matime}.

Let $I$ and $U$ be the set of all items and users, respectively \cite{ricciintro, huangnovel}. The user-item interactions form a bipartite graph $G = (U, I, E)$, where $E \subseteq U \times I$ is the set of observed interactions. 
The model updates user and item embeddings as follows -
\begin{equation}
e_{u}^{(k+1)} = \sum_{i \in \mathcal{I}(u)} 
\frac{1}{\sqrt{|\mathcal{I}(u)|}\sqrt{|\mathcal{U}(i)|}} \, e_{i}^{(k)},
\quad
e_{i}^{(k+1)} = \sum_{u \in \mathcal{U}(i)} 
\frac{1}{\sqrt{|\mathcal{U}(i)|}\sqrt{|\mathcal{I}(u)|}} \, e_{u}^{(k)}
\label{eq:propagation}
\end{equation}
where, $e_u^{(k)}$ and $e_i^{(k)}$ denote the embedding of user $u$ and item $i$ at the $k$-th layer, respectively \cite{liulearning}. 
$\mathcal{I}(u)$ is the set of items that are interacted by user $u$, and $\mathcal{U}(i)$ is the set of users that have interacted with item $i$. So, after $K$ layers, each user and item has multiple embeddings $\{ e_u^{(0)}, e_u^{(1)}, \ldots, e_u^{(K)} \}$ and $\{ e_i^{(0)}, e_i^{(1)}, \ldots, e_i^{(K)} \}$, respectively. All these embeddings are aggregated to obtain the final embedding for individual users and items.
\begin{equation}
e_u = \sum_{k=0}^{K} \alpha_k \, e_u^{(k)}, 
\qquad 
e_i = \sum_{k=0}^{K} \alpha_k \, e_i^{(k)}
\label{eq:aggregate}
\end{equation}
where $\alpha_k = \frac{1}{k + 1}$ denotes the significance of the $k$-th layer embedding. The model prediction for a user–item pair $(u, i)$ is obtained as follows -
\begin{equation}
p_{ui} = e_u^{\top} e_i
\label{eq;pred}
\end{equation}
where $p_{ui}$ is predicted preference score which signifies whether to recommend an item $i$ to a user $u$ or not \cite{liurepeat}. To learn the embeddings, the model adopts the Bayesian Personalized Ranking (BPR) loss~\cite{bib21, glara}, which is a pairwise ranking objective. For each user $u$, given an item $i \in \mathcal{I}(u)$ and another item $j \notin \mathcal{I}(u)$, the objective is to train the model such that the model learns to give higher preference to the items a user has interacted with, as opposed to the item the user has not interacted with \cite{pairwise, wuhybrid, huangadvanced}. 
\begin{equation}
\mathcal{L}_{\text{BPR}} =
- \sum_{u=1}^{|U|} \sum_{i \in \mathcal{I}(u)} 
\sum_{j \notin \mathcal{I}(u)}
\ln \sigma (p_{ui} - p_{uj})
+ \lambda \| E^{(0)} \|_2^2
\label{eq:bpr}
\end{equation}
where $\sigma(\cdot)$ denotes the sigmoid function. The predicted scores $p_{ui}$ and $p_{uj}$ are obtained using Equation \ref{eq;pred}, $E^{(0)}$ is the 0-th layer embeddings and $\lambda$ controls the strength of $L_2$ regularization \cite{bib6}. $E^{(0)}$ is updated during backpropagation. 
In this work, we utilize this model as baseline model to build our proposed Framework and designate it as $B$. We explain this in details in the following section.

\section{Proposed Method for Fair Recommendations}
\label{S:4}
As discussed previously, disparity in recommendation across individual users is an issue which needs to be addressed for developing a fair recommendation framework. We propose a dual-stage architecture which addresses this challenge. 
\begin{figure}[ht] 
  \centering
\includegraphics[width=\textwidth,height=0.6\textheight,keepaspectratio]{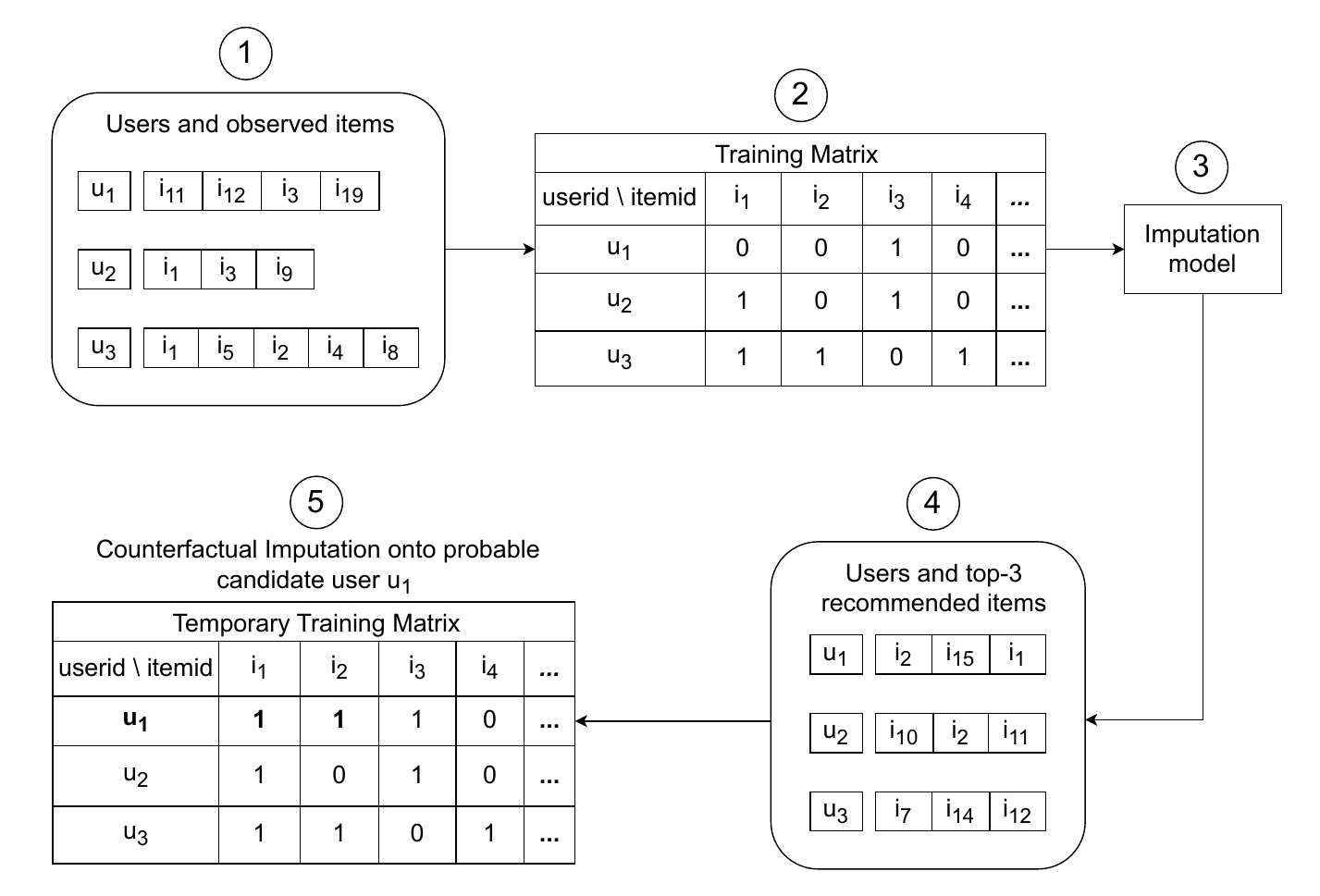}
  \caption{Imputation of items with highest preference scores for a probable candidate user to obtain modified training data.}
  \label{fig;cftimputation}
\end{figure}
In the first stage, a counterfactual technique is deployed to identify the suffering users. In the second stage, new interactions are introduced in favor of the identified users to alleviate their interaction profile.

In the first stage, a baseline model $B$ is utilized to compute the prediction score of each item, a user has interacted with in the training data. Main idea of this unconventional approach is to assess the effect of the learned embeddings of the model $B$ on the suffering users. Therefore, we evaluate the recommendation quality of the model $B$ by employing a commonly used metric namely, Normalized Discounted Cumulative Gain (NDCG). It can be noted that we compute NDCG by utilizing the information available in training data. In this work, this metric is also utilized as a disparity measure. We designate this metric by the notation $\mathrm{G}_{u}^{prior}$. The users getting low $\mathrm{G}_{u}^{prior}$ imply that model $B$ is biased towards them.

\begin{figure}[ht]
\centering 
\label{diagram}
\includegraphics[width=1.0\textwidth]{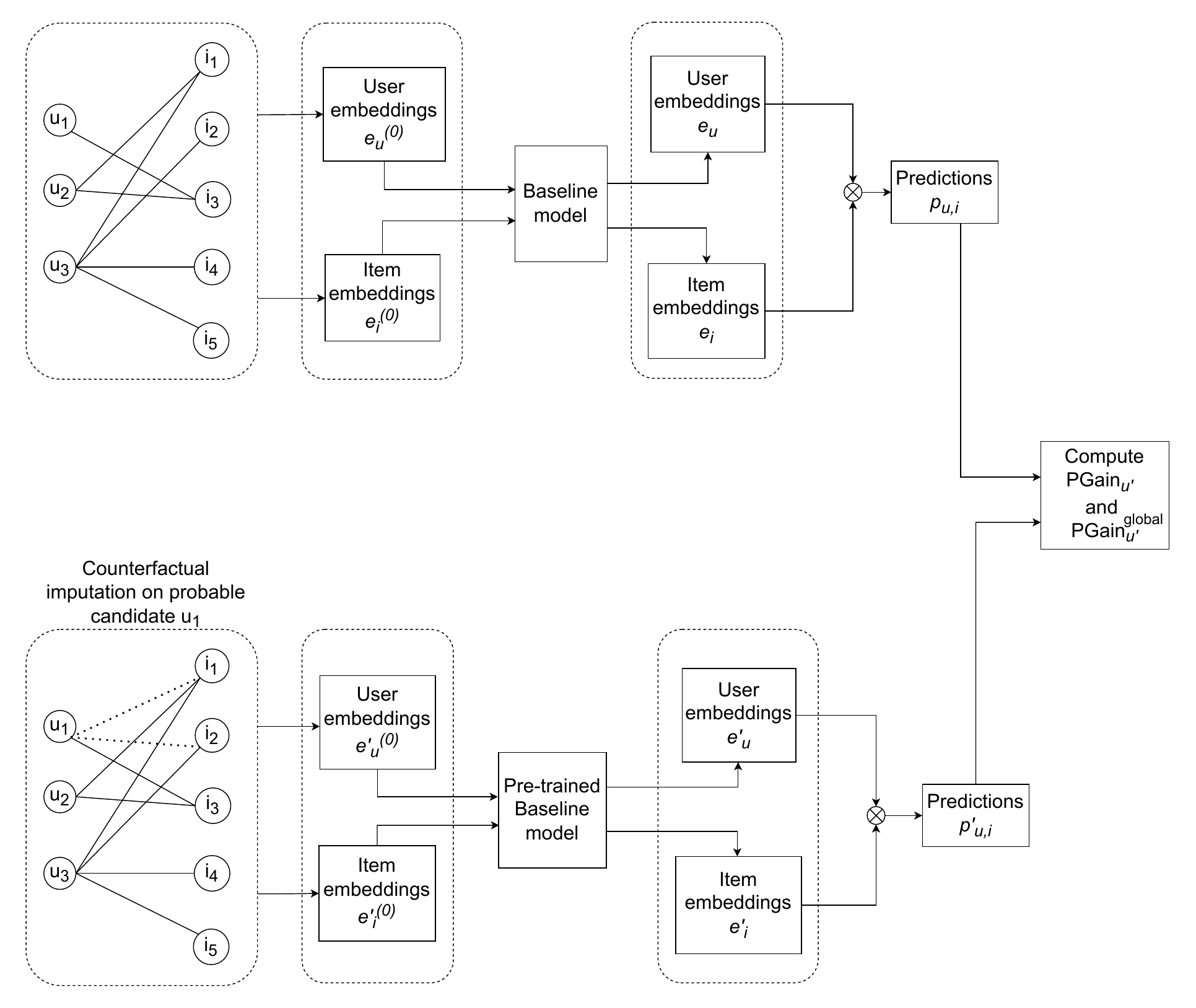}
\caption{Process of finding user candidates for final imputation consideration.}
\label{fig:ProposedFramework}
\end{figure}

As mentioned earlier, RSs are biased toward users with less number of interactions \cite{pham}. Therefore, we select a subset of users $U' \subset U$ as probable user candidates who are the suffering users and may benefit themselves (or the remaining users) by adding new interactions into the existing interaction graph. Hence, a user is chosen as a probable user candidate $u' \in U'$, if the user $u'$ either has a low $\mathrm{G}_{u}^{prior}$ or has few $|\mathcal{I}(u)|$.  

Having identified the probable user candidates, we temporarily add interactions to improve the user embeddings. For this purpose, we utilize a second recommendation model, which provides the recommended (unseen) items for these set of users. We designate the imputation model as $M$. These items are considered to be preferred by user $u'$ apart from the items they have previously interacted with. Hence, these items are counterfactually imputed on each of the probable user candidates $u'$. This process is shown using a toy example in Figure \ref{fig;cftimputation}. Step 1 consists of three users and the items each of them have interacted with, in their training data. In step 2, this information is transformed into a training matrix, $R$. In step 3, this matrix is fed into imputation model which recommends top-3 previously non-interacted items to the users. For this example, we assume u$_{\mathrm{1}}$ is a probable candidate. Hence, in step 5, a temporary training matrix is created where the top-3 recommended items for u$_{\mathrm{1}}$ are now considered as items this user has interacted with alongside the user's already seen items. 

Let $\mathcal{C}(u')$ be the set of items to be chosen for counterfactual imputation into the interaction graph of each user $u'$. The cardinality of $\mathcal{C}(u')$ is decided on the basis on this formula -
\begin{equation}
    |\mathcal{C}(u')| = \alpha \Bigg(\frac{|E|}{|U| \cdot |E'|}\Bigg)^{\beta}
\end{equation}
where, $\alpha$ regulates the overall magnitude of imputation, $E'\subset E$ is the set of observed interactions of probable user candidate $u'$, and $\beta$ controls how strongly the imputation count responds to user sparsity. The cardinality of $\mathcal{C}(u')$ is directly proportional to the global mean interactions for overall users and is inversely proportional to the total interactions of the individual candidate $u'$.  
We counterfactually impute $|\mathcal{C}(u')|$ interactions with the items $j \notin \mathcal{I}(u')$ having the highest prediction scores by $M$. The purpose is to observe the change in baseline model $B$'s predictions due to adding new interactions to $u'$. We obtain a temporary user–item interaction matrix $R'$ for each user $u'$. This adds temporary edges $(u', i)$ for $i \in \mathcal{C}(u')$ in the existing user–item bipartite graph. Subsequent propagation in LightGCN (Equation \ref{eq:propagation}) incorporates these new 
interactions, thereby refining both user and item embeddings \cite{cai}. So, after adding  $\mathcal{C}(u')$ the counterfactual embedding for the probable candidate $u'$ becomes -
\begin{equation}
e_{u'}^{(k+1)} = \sum_{i \in \mathcal{C}(u')  \cup \mathcal{I}(u')} 
\frac{1}{\sqrt{|\mathcal{C}(u') \cup \mathcal{I}(u)|}\sqrt{|\mathcal{U}(i)|}} \, e_{i}^{(k)}
\end{equation}
Each imputed item $i \in \mathcal{C}(u')$ gains a new edge from the user $u'$, which expands its set of neighboring users from $\mathcal{U}(i)$ to $\mathcal{U}(i) \cup \{u'\}$. Accordingly, the embedding update for all the items $i \in \mathcal{C}(u')$ at the $(k+1)$-th layer now becomes - 
\begin{equation}
e_{i}^{(k+1)} = 
\sum_{u \in \mathcal{U}(i) \cup \{u'\}} 
\frac{1}{\sqrt{|\mathcal{U}(i) \cup \{u'\}|}\sqrt{|\mathcal{I}(u)|}} \, e_{u}^{(k)}.
\end{equation}

The baseline model $B$ trains on this updated embeddings and minimizes BPR loss function (Equation \ref{eq:bpr}). The embedding for the various layers are aggregated using Equation \ref{eq:aggregate}. As BPR loss function encourages the model to rank $i$ higher than $j$, so, the baseline model's counterfactual predictions now becomes  $\tilde p_{ui}$ which is the modified preference score of user-item pairs after the counterfactual imputation on individual probable candidate $u'$. 
\begin{figure}[ht] 
  \centering
\includegraphics[width=\textwidth,height=0.6\textheight,keepaspectratio]{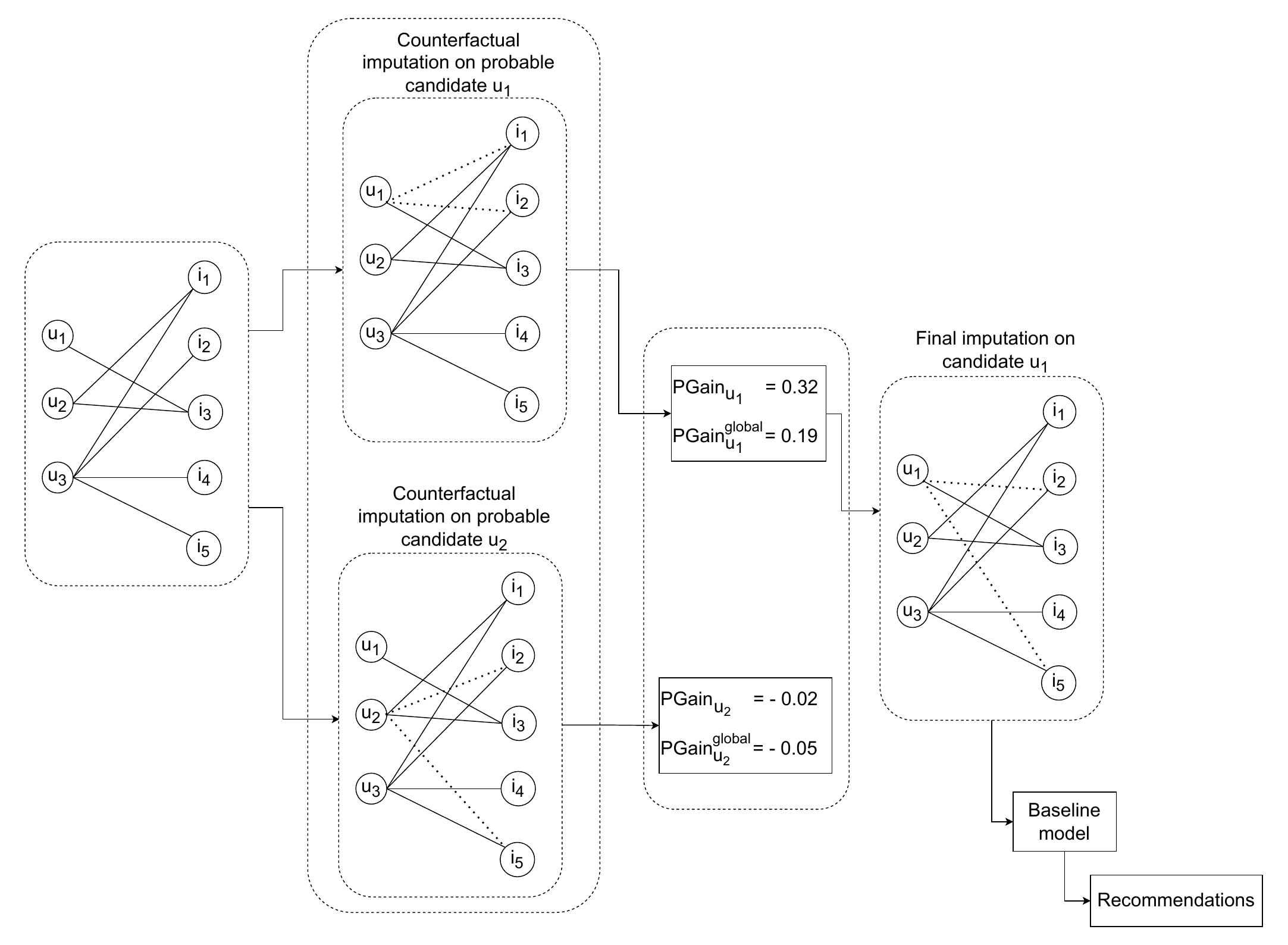}
  \caption{Imputation of items with highest preference scores for a candidate user to obtain modified training data.}
  \label{fig:3}
\end{figure}
A metric, Performance Gain (PGain) is introduced  which captures the model $B$'s performance for $u'$ prior and post counterfactual imputation. 
 \begin{equation}
    \mathrm{PGain}_{u'} = \mathrm{G}_{u'}^\mathrm{post}(R',Z) \;-\; \mathrm{G}_{u'}^\mathrm{prior}(R,P)
\end{equation}
$P$ is the prediction matrix obtained after raw training matrix $R$ on baseline model $B$. $Z$ is the matrix obtained after imputation and after model training. The users with negative $\mathrm{PGain}_{u'}$ mean that their individual performance has degraded after the imputation. And, the users with positive $\mathrm{PGain}_{u'}$ mean that their individual performance has improved after the imputation. As the embeddings get updated, this also influences the embeddings of other users. Thus, we compute the impact on the overall users due to $u'$'s counterfactual imputation. 
 \begin{equation}
    \mathrm{PGain}^\mathrm{global}_{u'} = \sum_{u=1}^{|U|}\mathrm{G}_{u}^\mathrm{post}(R',Z) \;-\; \sum_{u=1}^{|U|}\mathrm{G}_{u}^\mathrm{prior}(R,P)
\end{equation}
The users with negative $\mathrm{PGain}^\mathrm{global}_{u'}$ mean that the performance for overall users have degraded after the imputation. And, the users with positive $\mathrm{PGain}^\mathrm{global}_{u'}$ mean that the performance for overall users have improved after the imputation. After the model $B$ is optimized counterfactually on each of the probable user candidates $u'$, their corresponding $\mathrm{PGain}_{u'}$ and $\mathrm{PGain}^\mathrm{global}_{u'}$ is obtained. Additionally, $\mathcal{C}(u')$ now gets updated with the baseline model $B$'s top predicted (unseen) items for each user $u'$. This is demonstrated in Figure \ref{fig:ProposedFramework}. The users with either positive $\mathrm{PGain}_{u'}$ or positive $\mathrm{PGain}^\mathrm{global}_{u'}$ are selected as final user candidates $U^{f} \subseteq U'$. After this step, the second stage commences.

In the second stage, for all the users $u' \in U^{f}$, the items in the updated $\mathcal{C}(u')$ are added to the existing interaction graph. Now, this becomes the final training data, $R^{f}$. The entries of this matrix is obtained as follows - 
\begin{equation}
R^{f}_{ui} =
\begin{cases}
1, & \text{if }  {R}_{ui} =1 \ \text{or} \ i \in \mathcal{C}(u') \; \forall u' \in U^{f}, \\
0, & \text{otherwise.}
\end{cases}
\end{equation}
Baseline model $B$ now learns the embeddings of this updated training data $R^{f}$. Finally, Equation \ref{eq;pred} is evaluated to generate the final recommendations for all the users. 
Figure \ref{fig:3} pictorially demonstrates this framework. In Figure \ref{fig:3}, we assume there are two probable candidates u$_1$ and u$_2$. As neither $\mathrm{PGain}_{u_2}$ nor $\mathrm{PGain}^\mathrm{global}_{u_2}$ is positive, u$_2$ is dropped and only u$_1$ is considered as a candidate user. 

In this architecture, DualVAE \cite{bib22} is utilized as imputation model $M$.
Although, LightGCN and DualVAE models are utilized in the proposed architecture, the proposed architecture is model-agnostic. Any other recommendation model can be utilized in their place.

\section{Experimental Results and Analysis}
\label{S:5}

We conducted experiments on MovieLens 100K\footnote{https://grouplens.org/datasets/movielens/100k/} , MovieLens 1M\footnote{https://grouplens.org/datasets/movielens/1m/} and Amazon Beauty \footnote{https://cseweb.ucsd.edu/~jmcauley/datasets/amazon/links.html}, to show the performance of our proposed architecture \cite{fangart, li2022study}. For all the datasets, we retained users and items with minimum 10 interactions \cite{degree}. The statistics of the three datasets before and after this pre-processing step is shown in Table \ref{tab;statistics}.

\begin{table}[!htp]
\centering
\small
\caption{Statistics of the three datasets before and after pre-processing.}
\label{tab;statistics}
\begin{tabular}{p{1.6cm} *3{r}  *3{r}}
\midrule
 & \multicolumn{3}{c}{\textit{Before pre-processing}} 
 & \multicolumn{3}{c}{\textit{After pre-processing}}\\
\cmidrule(r){2-4} \cmidrule(r){5-7}
\textbf{Dataset} & \textbf{\#Users} & \textbf{\#Items} & \textbf{\#Ratings} 
& \textbf{\#Users} & \textbf{\#Items} & \textbf{\#Ratings}\\
\midrule
ML-100K & 943 & 1682 & 100,000 & 943 & 1152 & 97,953\\
A-Beauty & 1,210,271 & 249,274 & 2,023,070 & 1,340 & 733 & 28,798\\
ML-1M & 6,040 & 3,706 & 1,000,209 & 6,040 & 3,260 & 998,539\\
\midrule
\end{tabular}%
\end{table}

After the pre-processing step, we split the datasets into training and test sets based on timestamp. For each user, we retained the first 80\% of each user's interactions with the items as train set and the remaining 20\% as test set \cite{bib23, flexible, pugraph, music}. To evaluate the recommendation quality of proposed work, we used standard metrics Normalized Discounted Cumulative Gain (NDCG) \cite{ndcg1, huangadvanced}, and F1 for three values of k = 10, 15 and 20. For each user, all the unrated items in the pre-processed training data are considered as the items that can be recommended \cite{statistical, flexible}. We implemented the codes of proposed and related work in PyTorch \cite{pytorch} and TensorFlow \cite{tensorflow} environment. All the experiments are conducted on NVIDIA RTX A4000 system with driver version 553.35 and CUDA version 12.4.

\begin{table}[!htp]
\small
\centering
\caption{Recommendation performance of proposed and fairness models for overall users.}
\label{tab:fair_performance_overall}
\begin{tabular*}{\textwidth}{@{\extracolsep{\fill}} p{1.1cm} p{2.44cm} *{6}{r}}
\midrule
& & \multicolumn{6}{c}{\textit{Overall Users}}\\
\cmidrule{3-8}
\textbf{Dataset} & \textbf{Model} & \textbf{F1@10} & \textbf{N@10} & \textbf{F1@15} & \textbf{N@15} & \textbf{F1@20} & \textbf{N@20}\\
\midrule
\multirow{5}{*}{ML-100K} 
& Antidote \cite{bib9} & 0.0171 & 0.0269 & 0.0227 & 0.0289 & 0.0263 & 0.0305\\
& OCEF \cite{bib20} & 0.0355 & 0.0479 & 0.0397 & 0.0510 & 0.0445 & 0.0562\\
& DEE-KDE \cite{bib19} & 0.0655 & 0.0960 & 0.0745 & 0.0981 & 0.0802 & 0.1022\\
& ACFR-User$_{u}$ \cite{bib11} & 0.0338 & 0.0565 & 0.0386 & 0.0550 & 0.0422 & 0.0558\\
& Proposed & \textbf{0.1459} & \textbf{0.1993} & \textbf{0.1625} & \textbf{0.2053} & \textbf{0.1710} & \textbf{0.2139}\\
\midrule
\multirow{5}{*}{A-Beauty} 
& Antidote \cite{bib9} & 0.0082 & 0.0117 & 0.0083 & 0.0134 & 0.0088 & 0.0155\\
& OCEF \cite{bib20} & 0.0056 & 0.0065 & 0.0055 & 0.0076 & 0.0064 & 0.0096\\
& DEE-KDE \cite{bib19} & 0.0137 & 0.0175 & 0.0145 & 0.0215 & 0.0150 & 0.0251\\
& ACFR-User$_{u}$ \cite{bib11} & 0.0065 & 0.0078 & 0.0079 & 0.0106 & 0.0093 & 0.0136\\
& Proposed & \textbf{0.0509} & \textbf{0.0645} & \textbf{0.0509} & \textbf{0.0765} & \textbf{0.0492} & \textbf{0.0855}\\
\midrule
\multirow{5}{*}{ML-1M} 
& Antidote \cite{bib9} & 0.0488 & 0.0899 & 0.0595 & 0.0901 & 0.0676 & 0.0916\\
& OCEF \cite{bib20} & 0.0265 & 0.0415 & 0.0318 & 0.0434 & 0.0357 & 0.0460\\
& DEE-KDE \cite{bib19} & 0.0447 & 0.0858 & 0.0526 & 0.0832 & 0.0580 & 0.0828\\
& ACFR-User$_{u}$ \cite{bib11} & 0.0301 & 0.0664 & 0.0380 & 0.0645 & 0.0429 & 0.0637\\
& Proposed & \textbf{0.0861} & \textbf{0.1325} & \textbf{0.1032} & \textbf{0.1373} & \textbf{0.1129} & \textbf{0.1428}\\
\midrule
\end{tabular*}
\end{table}

The hyperparameter settings for baseline model $B$ is same as LGCN\footnote{https://github.com/gusye1234/pytorch-light-gcn}. Similarly, the hyperparameter settings for imputation model $M$ is same as DualVAE\footnote{https://github.com/georgeguo-cn/DualVAE}. Applying grid search, we obtain the combination of $\alpha$ = 10 and $\beta$ = 0.5 to produce optimal results. We compare our proposed architecture with existing models which address user unfairness. The models are Antidote \cite{bib9}, OCEF \cite{bib20}, DEE-KDE \cite{bib19} and ACFR-User$_{u}$\footnote{https://github.com/jasonshere/ACFR} \cite{bib11}.
Table \ref{tab:fair_performance_overall} compares the performance of these models with the proposed work for overall users. We observe in Table \ref{tab:fair_performance_overall} that proposed approach consistently outperforms the other methods. Apart from this, the results also follow the normal trend where the values of ndcg and F1 increase with the increasing value of k. We compare the performance of related work \cite{pmgcn} with the proposed work for the candidate users and demonstrate this in Table \ref{tab:fair_performance_candidate}. As mentioned in Section \ref{S:4}, the probable user candidates with either positive  $\mathrm{PGain}_{u'}$ or positive $\mathrm{PGain}^\mathrm{global}_{u'}$ are selected as candidate users for proposed approach. To maintain homogeneity in comparison, for each dataset we fix the number of candidate users same as proposed architecture. However, the individual candidate users vary for the different algorithms. We observe in Table \ref{tab:fair_performance_candidate} that DEE-KDE \cite{bib19} outperforms proposed approach for NDCG@10 in MovieLens 1M dataset. In all the other cases, proposed approach outperforms the other techniques.

\begin{table}[!htp]
\small
\centering
\caption{Recommendation performance of proposed and fairness models for candidate users.}
\label{tab:fair_performance_candidate}
\begin{tabular*}{\textwidth}{@{\extracolsep{\fill}} p{1.1cm} p{2.44cm} *{6}{r}}
\midrule
& & \multicolumn{6}{c}{\textit{Candidate Users}}\\
\cmidrule{3-8}
\textbf{Dataset} & \textbf{Model} & \textbf{F1@10} & \textbf{N@10} & \textbf{F1@15} & \textbf{N@15} & \textbf{F1@20} & \textbf{N@20}\\
\midrule
\multirow{5}{*}{ML-100K} 
& Antidote \cite{bib9} &
0.0172 & 0.0285 & 0.0179 & 0.0290 & 0.0153 & 0.0277\\
& OCEF \cite{bib20} &
0.0528 & 0.0644 & 0.0557 & 0.0774 & 0.0571 & 0.0847\\
& DEE-KDE \cite{bib19} &
0.0698 & 0.0750 & 0.0845 & 0.0914 & 0.0976 & 0.1102\\
& ACFR-User$_{u}$ \cite{bib11} &
0.0176 & 0.0301 & 0.0192 & 0.0305 & 0.0255 & 0.0369\\
& Proposed &
\textbf{0.1438} & \textbf{0.2163} & \textbf{0.1555} & \textbf{0.1992} & \textbf{0.1750} & \textbf{0.2072}\\
\midrule
\multirow{5}{*}{A-Beauty} 
& Antidote \cite{bib9} &
0.0016 & 0.0015 & 0.0013 & 0.0015 & 0.0020 & 0.0024\\
& OCEF \cite{bib20} &
0.0134 & 0.0164 & 0.0117 & 0.0179 & 0.0108 & 0.0201\\
& DEE-KDE \cite{bib19} &
0.0201 & 0.0260 & 0.0246 & 0.0349 & 0.0208 & 0.0362\\
& ACFR-User$_{u}$ \cite{bib11} &
0.0014 & 0.0015 & 0.0054 & 0.0065 & 0.0095 & 0.0122\\
& Proposed &
\textbf{0.0156} & \textbf{0.0240} & \textbf{0.0211} & \textbf{0.0301} & \textbf{0.0231} & \textbf{0.0351}\\
\midrule
\multirow{5}{*}{ML-1M} 
& Antidote \cite{bib9} &
0.0378 & 0.0444 & 0.0409 & 0.0521 & 0.0430 & 0.0599\\
& OCEF \cite{bib20} &
0.0220 & 0.0303 & 0.0265 & 0.0331 & 0.0314 & 0.0372\\
& DEE-KDE \cite{bib19} &
0.0375 & \textbf{0.0936} & 0.0474 & 0.0906 & 0.0533 & 0.0881\\
& ACFR-User$_{u}$ \cite{bib11} &
0.0258 & 0.0559 & 0.0321 & 0.0549 & 0.0390 & 0.0562\\
& Proposed &
\textbf{0.0704} & 0.0879 & \textbf{0.0855} & \textbf{0.0989} & \textbf{0.0936} & \textbf{0.1079}\\
\midrule
\end{tabular*}
\end{table}

As we built our architecture utilizing LGCN as baseline model, we conducted experiments and compare the performance of LGCN model with proposed work for candidate users (Table \ref{tab:baseline}). We observe that proposed approach outperforms the baseline model in all the three datasets. We also report the performance of LGCN and proposed approach for overall users. Interestingly,  for the overall users in MovieLens-1M dataset, there is minor improvement for all the cases, except for NDCG@15, where the performance remain same. This aligns with our objective as our goal is to improve the performance of candidate users without deteriorating the performance for the overall users. 

\begin{table}[!htp]
\small
\centering
\caption{Recommendation performance on three datasets for the baseline and proposed model on all users and candidate users.}
\label{tab:baseline}
\begin{tabular*}{\textwidth}{@{\extracolsep{\fill}} p{1.5cm} p{1.8cm} *{6}{r}}
\midrule
& & \textbf{F1@10} & \textbf{N@10} & \textbf{F1@15} & \textbf{N@15} & \textbf{F1@20} & \textbf{N@20}\\
\cmidrule{3-8}
\textbf{Dataset} & \textbf{Method} & \multicolumn{6}{c}{\textit{Overall Users}} \\
\midrule

\multirow{6}{*}{ML-100K} 
& LGCN~\cite{bib6} & 0.1342 & 0.1843 & 0.1505 & 0.1896 & 0.1625 & 0.2002\\
& Proposed & \textbf{0.1459} & \textbf{0.1993} & \textbf{0.1625} & \textbf{0.2053} & \textbf{0.1710} & \textbf{0.2139}\\
\cmidrule{2-8}

& & \multicolumn{6}{c}{\textit{Candidate Users}} \\
\cmidrule{2-8}
& LGCN~\cite{bib6} &
0.1099 & 0.1629 & 0.1413 & 0.1663 & 0.1609 & 0.1774\\
& Proposed &
\textbf{0.1438} & \textbf{0.2163} & \textbf{0.1555} & \textbf{0.1992} & \textbf{0.1750} & \textbf{0.2072}\\
\midrule

& & \multicolumn{6}{c}{\textit{Overall Users}} \\
\cmidrule{2-8}
\multirow{6}{*}{A-Beauty} 
& LGCN~\cite{bib6} & 0.0205 & 0.0262 & 0.0211 & 0.0316 & 0.0220 & 0.0369\\
& Proposed & \textbf{0.0509} & \textbf{0.0645} & \textbf{0.0509} & \textbf{0.0765} & \textbf{0.0492} & \textbf{0.0855}\\

\cmidrule{2-8}
& & \multicolumn{6}{c}{\textit{Candidate Users}} \\
\cmidrule{2-8}
& LGCN~\cite{bib6} &
0.0079 & 0.0084 & 0.0074 & 0.0094 & 0.0104 & 0.0127\\
& Proposed &
\textbf{0.0156} & \textbf{0.0240} & \textbf{0.0211} & \textbf{0.0301} & \textbf{0.0231} & \textbf{0.0351}\\
\midrule

& & \multicolumn{6}{c}{\textit{Overall Users}} \\
\cmidrule{2-8}
\multirow{6}{*}{ML-1M} 
& LGCN~\cite{bib6} & 0.0860 & 0.1324 & 0.1027 & \textbf{0.1373} & 0.1124 & 0.1427\\
& Proposed & \textbf{0.0861} & \textbf{0.1325} & \textbf{0.1032} & \textbf{0.1373} & \textbf{0.1129} & \textbf{0.1428}\\

\cmidrule{2-8}
& & \multicolumn{6}{c}{\textit{Candidate Users}} \\
\cmidrule{2-8}
& LGCN~\cite{bib6} &
0.0691 & 0.0835 & 0.0854 & 0.0942 & 0.0913 & 0.1055\\
& Proposed &
\textbf{0.0704} & \textbf{0.0879} & \textbf{0.0855} & \textbf{0.0989} & \textbf{0.0936} & \textbf{0.1079}\\
\midrule
\end{tabular*}
\end{table}  

\small
\begin{table}[!htp]
\centering
\caption{The recommendation performance of proposed work on MovieLens 100K with and without the counterfactual segment, for k = 10.}
\label{tab:ablation10}
\begin{tabular}{c p{1.2cm} p{3.8cm} *{4}{r}}
\midrule
 & & & \textbf{\#Users} 
& \textbf{Before} & \textbf{After} & \textbf{PGain} \\
\cmidrule{4-7}
\textbf{k} & \textbf{Metric} & \textbf{Proposed Work} & \multicolumn{4}{c}{\textit{Candidate Users}}\\
\midrule
\multirow{9}{*}{10} & \multirow{2}{*}{F1} 
 & without counterfactual & 372 
 & 0.1346 & 0.1210 & -0.0136\\
& & with counterfactual & 27 
 & 0.1099 & 0.1438 & \textbf{0.0339}\\
\cmidrule(l){2-7}
& \multirow{2}{*}{NDCG} 
 & without counterfactual & 372 
 & 0.1600 & 0.1395 & -0.0205\\
& & with counterfactual & 27 
 & 0.1629 & 0.2163 & \textbf{0.0534}\\
\cmidrule(l){2-7}
& & & \multicolumn{4}{c}{\textit{Remaining Users}}\\
\cmidrule(l){2-7}
& \multirow{2}{*}{F1} 
 & without counterfactual & 571 
 & 0.1219 & 0.1011 & -0.0208\\
& & with counterfactual & 916 
 & 0.1349 & 0.1459 & \textbf{0.011}\\
\cmidrule(l){2-7}
& \multirow{2}{*}{NDCG} 
 & without counterfactual & 571 
 & 0.2015 & 0.1752 & -0.0263\\
& & with counterfactual & 916 
 & 0.1849 & 0.1988 & \textbf{0.0139}\\
 \midrule
\end{tabular}%
\end{table}

\small
\begin{table}[htp]
\centering
\caption{The recommendation performance of proposed work on MovieLens 100K with and without the counterfactual segment, for k = 15.}
\label{tab:ablation15}
\begin{tabular}{c p{1.2cm} p{3.8cm} *{4}{r}}
\midrule
 & & & \textbf{\#Users} 
& \textbf{Before} & \textbf{After} & \textbf{PGain} \\
\cmidrule{4-7}
\textbf{k} & \textbf{Metric} & \textbf{Method} & \multicolumn{4}{c}{\textit{Candidate Users}}\\
\midrule
\multirow{9}{*}{15} & \multirow{2}{*}{F1} 
 & without counterfactual & 372 
 & 0.1411 & 0.1302 & -0.0109\\
& & with counterfactual & 27 
 & 0.1413 & 0.1555 & \textbf{0.0142}\\
\cmidrule(l){2-7}
& \multirow{2}{*}{NDCG} 
 & without counterfactual & 372 
 & 0.1785 & 0.1585 & -0.02\\
& & with counterfactual & 27 
 & 0.1663 & 0.1992 & \textbf{0.0329}\\
\cmidrule(l){2-7}
& & & \multicolumn{4}{c}{\textit{Remaining Users}}\\
\cmidrule(l){2-7}
& \multirow{2}{*}{F1} 
 & without counterfactual & 571 
 & 0.1426 & 0.1215 & -0.0211\\
& & with counterfactual & 916 
 & 0.1508 & 0.1626 & \textbf{0.0118}\\
\cmidrule(l){2-7}
& \multirow{2}{*}{NDCG} 
 & without counterfactual & 571 
 & 0.1982 & 0.1730 & -0.0252\\
& & with counterfactual & 916 
 & 0.1903 & 0.2054 & \textbf{0.0151}\\
 \midrule
\end{tabular}%
\end{table}

\small
\begin{table}[htp]
\centering
\caption{The recommendation performance of proposed work on MovieLens 100K with and without the counterfactual segment, for k = 20.}
\label{tab:ablation20}
\begin{tabular}{c p{1.2cm} p{3.8cm} *{4}{r}}
\midrule
 & & & \textbf{\#Users} 
& \textbf{Before} & \textbf{After} & \textbf{PGain} \\
\cmidrule{4-7}
\textbf{k} & \textbf{Metric} & \textbf{Method} & \multicolumn{4}{c}{\textit{Candidate Users}}\\
\midrule
\multirow{9}{*}{20} & \multirow{2}{*}{F1} 
 & without counterfactual & 372 
 & 0.1450 & 0.1299 & -0.0151\\
& & with counterfactual & 27 
 & 0.1609 & 0.1750 & \textbf{0.0141}\\
\cmidrule(l){2-7}
& \multirow{2}{*}{NDCG} 
 & without counterfactual & 372 
 & 0.1999 & 0.1739 & -0.0260\\
& & with counterfactual & 27 
 & 0.1774 & 0.2072 & \textbf{0.0298}\\
\cmidrule(l){2-7}
& & & \multicolumn{4}{c}{\textit{Remaining Users}}\\
\cmidrule(l){2-7}
& \multirow{2}{*}{F1} 
 & without counterfactual & 571 
 & 0.1582 & 0.1380 & -0.0202\\
& & with counterfactual & 916 
 & 0.1624 & 0.1708 & \textbf{0.0084}\\
\cmidrule(l){2-7}
& \multirow{2}{*}{NDCG} 
 & without counterfactual & 571 
 & 0.2017 & 0.1776 & -0.0241\\
& & with counterfactual & 916 
 & 0.2008 & 0.2141 & \textbf{0.0133}\\
 \midrule
\end{tabular}%
\end{table}

We perform ablation study \cite{pmgcn} on ML-100K dataset, to justify the utilization of counterfactual approach before the actual imputation in the proposed approach. We report this by comparing the performance of proposed architecture with (without) the counterfactual imputation segment. We compute PGain for each of the candidate users and average it to produce the results shown in the Table \ref{tab:ablation10}, Table \ref{tab:ablation15} and Table \ref{tab:ablation20} for the different values of k. In all these three tables, we observe that without the counterfactual segment the number of candidate users is 372 which is very large as compared to 27. Although the candidates are far more in the proposed approach without counterfactual segment, after performing the final imputation, PGain drops. For the proposed work with counterfactual segment, however, we observe consistent improvement in PGain. Apart from the candidate users, we also track the performance of the remaining users. Even for the remaining users, the proposed approach with counterfactual segment shows improvement in PGain. On the contrary, there is drop in PGain for the proposed approach without counterfactual segment for the remaining users. This trend is a clear indication of the efficacy of the proposed approach.

\FloatBarrier
\section{Conclusion }\label{sec13}

Traditional recommender systems tend to be biased against users with limited interaction history. Therefore, these  users get poorer recommendations than the rest. In this paper, we proposed a framework for addressing individual user unfairness in recommender systems. The proposed methodology identifies the victim  users and subsequently, a counterfactual technique is adopted to mitigate the bias towards the victim users. We counterfactually introduce new interactions to the victim users once at a time and analyze the imputation effect. Finally, interactions which provide gain in recommendation quality is incorporated in the training dataset. We conducted extensive experiments on 
three datasets and demonstrated the superiority of the proposed work. This work can be extended by exploring the impact of other performance metrics while incorporating interactions.

\section*{Declarations}
\textbf{Conflicts of interests} The authors declare that they have no competing
interests.

\bibliography{sn-bibliography}

\end{document}